\documentclass[a4paper,12pt]{article}
\usepackage{amssymb}
\usepackage{amsmath}
\usepackage{mathtools}
\usepackage{slashed}
\usepackage{color}
\usepackage{indentfirst}
\usepackage{graphicx}
\usepackage{csquotes} 
\usepackage{caption}
\usepackage{cite}
\renewcommand{\thefootnote}{\fnsymbol{footnote}}

\begin{document}

\title{
\begin{flushright}
\begin{minipage}{0.2\linewidth}
\normalsize
EPHOU-20-004\\
KEK-TH-2206 \\*[50pt]
\end{minipage}
\end{flushright}
{\Large \bf 
Challenge for spontaneous CP violation\\ 
in Type IIB orientifolds with fluxes
\\*[20pt]}}

\author{Tatsuo Kobayashi$^{a}$\footnote{
E-mail address: kobayashi@particle.sci.hokudai.ac.jp
}
\ and\
Hajime~Otsuka$^{b}$\footnote{
E-mail address: hotsuka@post.kek.jp
}\\*[20pt]
$^a${\it \normalsize 
Department of Physics, Hokkaido University, Sapporo 060-0810, Japan} \\
$^b${\it \normalsize 
KEK Theory Center, Institute of Particle and Nuclear Studies, KEK,}\\
{\it \normalsize 1-1 Oho, Tsukuba, Ibaraki 305-0801, Japan}}
\maketitle

\date{
\centerline{\small \bf Abstract}
\begin{minipage}{0.9\linewidth}
\medskip 
\medskip 
\small
We systematically investigate possibilities of realizing the spontaneous CP violation in Type IIB flux compactifications on toroidal orientifolds. Our detailed analysis leads to the presence of flat directions at degenerate CP-breaking and -conserving vacua for a generic choice of three-form fluxes, indicating that flux compactifications are not sufficient to realize the spontaneous CP violation. Furthermore, the four-dimensional CP can be embedded into the duality symmetries, 
namely modular symmetries for a particular choice of fluxes. 
\end{minipage}
}

\renewcommand{\thefootnote}{\arabic{footnote}}
\setcounter{footnote}{0}
\thispagestyle{empty}
\clearpage
\addtocounter{page}{-1}

\tableofcontents

\section{Introduction}
\label{sec:intro}

Understanding the origin of CP violation in the Standard Model (SM) is of particular interest not only to explain the structure of Yukawa couplings but also to search for new physics beyond the SM through experiments on the neutron electric dipole moment~\cite{Abel:2020gbr}, $K$ meson decays (probed by the KOTO experiment at the J-PARC and the NA62 experiment at CERN), and observations of the excess of baryons over antibaryons in the universe. 
It is interesting to ask the origin of CP violation in string theory.

As pointed out in Refs.~\cite{Dine:1992ya,Choi:1992xp}, CP is regarded as a 
discrete gauge symmetry in a class of $E_8\times E_8$ and $SO(32)$ heterotic string theories. Calabi-Yau (CY) compactifications lead to the spontaneous breaking of CP 
by the dynamics of massless scalar fields, called moduli fields~\cite{Strominger:1985it}. 
Here, the four-dimensional (4D) CP is embedded into the 10D proper Lorentz 
symmetry, indicating that the orientations of 4D spacetime and 
6D CY threefolds are reversed simultaneously. 
In the heterotic string context, axionic components of 
axio-dilaton and K\"ahler moduli have an odd parity under CP transformation~\cite{Strominger:1985it} 
and then, the 4D CP can be spontaneously broken by their vacuum 
expectation values (vevs). However, it is difficult to achieve the stabilization 
of the moduli fields on CY threefolds as well as toroidal orbifolds in 
a controlled way, due to the sizable backreactions from their stabilization.\footnote{It might be possible to break CP through vevs of charged scalar fields under the 
anomalous U(1) symmetry~\cite{Giedt:2002ns}.　
For the CP violation in the orbifold context, see, Refs.~\cite{Kobayashi:1994ks,Nilles:2018wex}.}

The purpose of this paper is to systematically investigate possibilities of realizing the spontaneous CP violation 
in Type IIB string theory on toroidal orientifolds, where the moduli stabilization as well as the calculation of Yukawa couplings is performed in a controlled way. 
Similar to the heterotic string theory, CP-odd axionic fields break the 4D CP spontaneously and their vevs induce the nonvanishing Cabbibo-Kobayashi-Maskawa (CKM) phase. 
Since it is possible to stabilize these axionic fields by three-form fluxes 
in the low-energy effective action of Type IIB string theory, we search for CP-breaking minima in Type IIB flux compactifications on several toroidal orientifolds. 
Starting from CP-invariant potentials, 
we find that flat directions exist at degenerate CP-breaking and -conserving 
vacua for a generic choice of three-form fluxes. 
To realize the spontaneous CP violation, it is required to resolve this degeneracy by 
other sources. 
Furthermore, 4D CP can be embedded into the duality symmetries, namely 
modular symmetries for a particular choice of fluxes. 
The approach to unify CP and modular symmetries is recently developed in Refs.~\cite{Dent:2001cc,Baur:2019kwi,Baur:2019iai,Kobayashi:2019uyt} and 
the relationship between the strong CP and CKM phases are also pointed out in Ref.~\cite{Kobayashi:2020oji}. 
Indeed, axions associated with the complex structure moduli of tori play a crucial role in such a “generalized CP" context \cite{Feruglio:2012cw,Holthausen:2012dk,Chen:2014tpa,Novichkov:2019sqv}.

This paper is organized as follows. 
In Section~\ref{sec:2}, we briefly review the origin of 4D CP in 
the effective action of Type II string theory with intersecting/magnetized 
D-branes. 
In Section~\ref{sec:3}, we discuss the CP-invariant low-energy effective action of Type IIB string theory on toroidal orientifolds. 
Three-form fluxes cause both the stabilization of axionic fields and spontaneous breaking of CP. 
It turns out that massless fields generically appear in the flux vacua on 
several toroidal orientifolds. 
Finally, we conclude in Section~\ref{sec:con}.

\section{Origin of CP in intersecting/magnetized D-brane models}
\label{sec:2}

In this section, we review the low-energy effective action of magnetized D-branes in Type IIB string theory on toroidal orientifolds with a special emphasis on 4D CP. 
It is applicable to Type IIA intersecting D-brane system 
by T-dualizing the Type IIB magnetized D-brane models.

Let us consider magnetized D7-branes wrapping the tori $(T^2)_i\times (T^2)_j$ with $i\neq j$. When $U(1)$ gauge boson on D7-branes takes a non-trivial vev on the torus, the internal space carries  background gauge flux $F^i$ on each torus $(T^2)_i$,
\begin{align}
    \frac{m^i}{l_s^2}\int_{(T^2)_i} F^i =n^i,
\end{align}
where $m^i$ and $n^i$ correspond to the wrapping number of magnetized D7-branes and the flux quanta, respectively. $l_s=2\pi \sqrt{\alpha^\prime}$ denotes the string length. 
Such gauge fluxes are relevant to determine 4D CP as discussed later. 

After compactifying on tori, the 4D $\theta$ term in the effective action of D7-branes 
is determined by the following gauge kinetic function~\cite{Lust:2004cx,Lust:2004fi,Lust:2004dn}
\begin{align}
    f_{{\rm D}7} =|m^im^j|\left( T^k -\frac{n^in^j}{m^im^j}S \right)
\label{eq:gaugekin}
\end{align}
with $i\neq j \neq k$. Here, we denote the K\"ahler moduli $T^i$ and the axio-dilaton $S$, respectively. 
Remarkably, the CP-odd axionic components of these moduli fields 
(Ramond-Ramond fields) ${\rm Re}(T^i)=\int_{(T^2)_j\times (T^2)_k} C_4$ and ${\rm Re}(S)=C_0$ induce the 4D $\theta$ term.

The CKM phase in Yukawa couplings of chiral zero-modes on magnetized D7-branes is also determined by one of the closed string moduli, namely complex structure moduli $\tau_i$ associated with tori $(T^2)_i$. 
Indeed, the presence of magnetic fluxes 
induces the multiple number of chiral zero-modes with non-trivial moduli-dependent Yukawa couplings. 
From the calculation of Yukawa couplings of chiral zero-modes in the conformal field theory as well as the field theoretical approach, it is known that holomorphic Yukawa couplings on each 2-torus $(T^2)_j$ inside the 4-cycle wrapped by magnetized D7-branes are given by the Jacobi theta function~\cite{Cremades:2003qj,Cvetic:2003ch,Abel:2003vv,Cremades:2004wa}
\begin{align}
\vartheta
\begin{bmatrix}
c\\
0
\end{bmatrix}
(0, \tau_j)
\equiv 
\sum_{l\in \mathbb{Z}}
e^{\pi i (c+l)^2\tau_j}
.
\end{align}

For illustrative purposes, let us suppose that $U(N)$ gauge symmetry realized in $N$ stacks 
of D7-branes is broken by magnetic fluxes to $U(N_a)\times U(N_b)\times U(N_c)$ with $N=N_a+N_b+N_c$. 
Then, bifundamental zero-modes $(N_\alpha, \bar{N_\beta})$, $\alpha, \beta =a,b,c$ have $I_{\alpha \beta}^i =n_\alpha^i/m_\alpha^i-n_\beta^i/m_\beta^i$ number of zero-modes on each torus $(T^2)_i$. 
Holomorphic Yukawa couplings of such zero-modes are provided by
\begin{align}
Y_{pqs} 
&= 
\vartheta
\begin{bmatrix}
-\frac{1}{I_{ab}^i}\left(\frac{q}{I_{ca}^i}+\frac{s}{I_{bc}^i}\right) \\ 0
\end{bmatrix}
\left(0,\tau_i \left|I_{ab}^iI_{bc}^iI_{ca}^i\right|\right)
,
\label{eq:Yukawa}
\end{align}
up to the normalization factor. 
Here $q=0,1,\cdots, |I_{ca}^i|-1$ and  $s=0,1,\cdots, |I_{bc}^i|-1$ label the number of index (generation number) and we employ $p=s-q$ mod $I_{ab}$~\cite{Cremades:2004wa}. 

In this way, axionic components of the complex structure moduli determine 
the magnitude of CKM phase, namely the Jarlskog determinant. 
(See for explicit examples, e.g. \cite{Kobayashi:2016qag}.)
The moduli-dependent Yukawa couplings as well as the gauge kinetic function in 
Type IIA intersecting D6-branes have the analogous form by T-dualizing the Type 
IIB magnetized D7-brane models~\cite{Cremades:2003qj}. 
For Type IIB string theory with D5/D9-branes, the gauge kinetic function 
as well as Yukawa couplings are also dependent on the K\"ahler moduli and 
complex structure moduli, respectively. 
It is remarkable that the complex structure moduli control 
not only the 4D CP but also the flavor symmetry in Type IIB 
string theory \footnote{See for a relation between the flavor symmetry and the modular symmetry of $\tau$ 
in type IIB string theory, e.g. \cite{Kobayashi:2018rad,Kobayashi:2018bff}.
}.

\section{Four-dimensional CP in flux compactifications}
\label{sec:3}

So far, we have discussed the origin of 4D CP phases in the effective action of 
magnetized D-branes. 
This section is devoted to explore the spontaneous CP violation 
on the basis of the effective action of moduli fields. 
We draw  general conclusions about the spontaneous CP violation in 
Type IIB toroidal orientifolds with fluxes. 

\subsection{CP-invariant moduli potential}
\label{sec:2_1}

We begin with the simplest $T^6/\mathbb{Z}_2$ orientifold, following the 
convention of Ref.~\cite{Kachru:2002he}. It is straightforward to extend the setup to 
$T^6/(\mathbb{Z}_2\times \mathbb{Z}_2^\prime)$ and other orbifolds, 
to be discussed later.

As argued in Ref. \cite{Dine:1992ya,Choi:1992xp}, the 4D CP is 
regarded as the higher-dimensional proper Lorentz symmetry, 
in particular 10D proper Lorentz symmetry in string theory. 
When the extra 6D space consists of the factorizable $T^6$ 
subject to $\mathbb{Z}_2$ identification, namely $\Pi_{i=1}^3(T^2)_i/\mathbb{Z}_2$ 
with coordinates $z_i$, 
10D proper Lorentz symmetry is defined by the 4D CP and 
6D transformations $z_i\rightarrow z_i^\ast$ or $z_i\rightarrow -z_i^\ast$ leading 
to the negative determinant in the transformation of 6D space. 
In this paper, we restrict ourselves to $z_i\rightarrow -z_i^\ast$ transformations, 
allowing us to consider ${\rm Im}\tau_i>0$ in the complex structure moduli spaces. According to $z_i\rightarrow -z_i^\ast$, $\tau_i$ transform as
\begin{align}
    \tau_i \rightarrow -\bar{\tau}_i
\end{align}
and at the same time, the axio-dilaton $S=C_0 +i e^{-\phi}$ 
also transforms
\begin{align}
    S \rightarrow -\bar{S}.
\end{align}
Note that the real part of the axio-dilaton is the Ramond-Ramond 
axion. 

Let us discuss CP invariance of the moduli effective action on the basis 
of $\Pi_{i=1}^3(T^2)_i/\mathbb{Z}_2$ orientifold. 
Under the above transformations, the moduli K\"ahler potential
\begin{align}
    K = -\ln (-i(S -\Bar{S})) -2\ln {\cal V} 
    -\ln(i(\tau_1-\Bar{\tau}_1)(\tau_2-\Bar{\tau}_2)(\tau_3-\Bar{\tau}_3))     
    \label{eq:Keff}
\end{align}
is invariant.
Here we denote ${\cal V}$ the volume of tori and employ the reduced Planck mass unit $M_{\rm Pl}=1$. 
On the other hand, the superpotential is induced by an 
existence of background three-form flux $G_3$~\cite{Gukov:1999ya}
\begin{align}
    W = \frac{1}{l_s^2}\int \Omega \wedge G_3,
    \label{eq:Weff}
\end{align}
with $l_s = 2\pi \sqrt{\alpha^\prime}$.\footnote{The normalization 
factor to match with the dimensional reduction of Type IIB supergravity action 
is irrelevant to the following discussion and therefore we omit such a factor for 
simplicity.} Now a holomorphic three-form 
$\Omega$ 
\begin{align}
    \Omega &= dz_1 \wedge dz_2 \wedge dz_3,    
    \label{eq:Omega}
\end{align}
transforms under the CP transformation as
\begin{align}
    \Omega \rightarrow -\bar{\Omega}.
\end{align}
The three-form flux $G_3$ is defined by a linear combination of 
Ramond-Ramond (RR) $F_3$ and Neveu-Schwarz (NS) three-forms $H_3$, namely 
$G_3 =F_3 -S H_3$, both of which are expanded on the basis of $H^3(T^6,\mathbb{Z})$,
\begin{align}
\frac{1}{l_s^2}F_3 &= a^0 \alpha_0 +a^{i}\alpha_{i} +b_{i}\beta^{i} +b_0 \beta^0,
\nonumber\\
\frac{1}{l_s^2}H_3 &= c^0 \alpha_0 +c^{i}\alpha_{i} +d_{i}\beta^{i} +d_0 \beta^0,
\label{eq:F3H3_T6Z2}
\end{align}
with
\begin{align}
    \alpha_0 &= dx^1 \wedge dx^2 \wedge dx^3, \qquad
    \alpha_{1} = dy^1 \wedge dx^2 \wedge dx^3,
\nonumber\\
    \alpha_2 &= dy^2 \wedge dx^3 \wedge dx^1, \qquad
    \alpha_3 = dy^3 \wedge dx^1 \wedge dx^2,
\nonumber\\
    \beta^0 &= dy^1 \wedge dy^2 \wedge dy^3, \qquad
    \beta^1 = -dx^1 \wedge dy^2 \wedge dy^3,
\nonumber\\
    \beta^2 &= -dx^2 \wedge dy^3 \wedge dy^1, \qquad
    \beta^3 = -dx^3 \wedge dy^1 \wedge dy^2,
\label{eq:basis}
\end{align}
satisfying $\int_{T^6}\alpha_I \wedge \beta^J =\delta^{J}_I$. 
Note that flux quanta $\{a^{0,1,2,3},b_{0,1,2,3},c^{0,1,2,3},d_{0,1,2,3}\}$ 
are even and/or odd integers depending on the existence of exotic ${\rm O3}'$-planes~\cite{Hanany:2000fq,Witten:1997bs} 
and they are 
constrained by the tadpole cancellation condition
\begin{align}
    n_{\rm flux}= \frac{1}{l_s^2}\int H_3\wedge F_3 = c^0b_0 -d_0a^0 +\sum_i (c^ib_i -d_ia^i) =32 -2n_{\rm D3}-n_{{\rm O3}'} \leq 32,
    \label{eq:nD3}
\end{align}
where $n_{\rm D3}$ and $n_{{\rm O3}'}$ are the number of D3-branes and exotic ${\rm O3}'$-planes, respectively. Here, anti-D3 brane contributions are not taken into account to preserve the supersymmetry in our system. 

Since the coordinate transformations $z_i \rightarrow -z_i^\ast$ 
correspond to $x_i \rightarrow -x_i$ and $y_i \rightarrow y_i$, 
these three-form bases also transform as
\begin{align}
    &\alpha_0 \rightarrow -\alpha_0,\qquad 
    \beta^0 \rightarrow \beta^0,\qquad 
    \alpha_i \rightarrow \alpha_i, \qquad
    \beta^i \rightarrow -\beta^i, \qquad (i=1,2,3).
\end{align}

To be invariant under the CP transformation, the 
flux-induced superpotential should be transformed as
\begin{align}
    W \rightarrow e^{i\gamma} \bar{W},
\end{align}
with $\gamma = 0$ or $\pi$ (mod $2\pi$), 
because $F_3$ and $H_3$ are quantized as real numbers. 
Recall that only complex quantity is $S$ in $G_3$, 
$\gamma \neq 0, \pi$ (mod $2\pi$) is contradicted 
with the real three-forms $F_3$ and $H_3$. 
Hence, the three-form fluxes obey
\begin{align}
    G_3 \rightarrow -e^{i\gamma} \bar{G}_3.
\end{align}
It restricts ourselves to the following two patterns of 
RR and NSNS three-forms,
\begin{itemize}
    \item $\gamma=0$ (mod $2\pi$)
\begin{align}
\frac{1}{l_s^2}F_3 &= a^0 \alpha_0 +b_{i}\beta^{i},
\nonumber\\
\frac{1}{l_s^2}H_3 &= c^{i}\alpha_{i} +d_0 \beta^0,
\label{eq:F3H3odd}
\end{align}
leading to $G_3 \rightarrow -F_3 +\bar{S}H_3 =-\bar{G}_3$.
\item $\gamma=\pi$ (mod $2\pi$)
\begin{align}
\frac{1}{l_s^2}F_3 &= a^{i}\alpha_{i} +b_0 \beta^0,
\nonumber\\
\frac{1}{l_s^2}H_3 &= c^0 \alpha_0 +d_{i}\beta^{i},
\label{eq:F3H3even}
\end{align}
\end{itemize}
leading to $G_3 \rightarrow F_3 -\bar{S}H_3 =\bar{G}_3$. 
Other choices are impossible to obtain the nonvanishing 
CP-invariant superpotential because of the real three-forms $F_3$ and $H_3$.

Then, it results in the two classes of the superpotential:
\begin{itemize}
    \item $\gamma=0$ (mod $2\pi$)
\begin{align}
    W &= a^0 \tau_1\tau_2\tau_3  +c^1 S \tau_2\tau_3
    +c^2 S\tau_1\tau_3 +c^3 S\tau_1\tau_2
 - \sum_{i=1}^3 b_i \tau_i + d_0 S,
\label{eq:Wodd}
\end{align}
    \item $\gamma=\pi$ (mod $2\pi$)
\begin{align}
    W &= -c^0 S\tau_1\tau_2\tau_3 - a^1 \tau_2\tau_3
    - a^2\tau_1\tau_3 - a^3\tau_1\tau_2
+ \sum_{i=1}^3 d_i S \tau_i - b_0.
\label{eq:Weven}
\end{align}
\end{itemize}

Note that the above statement also holds for $\Pi_{i=1}^3(T^2)_i/(\mathbb{Z}_2\times \mathbb{Z}_2^\prime)$ 
orientifolds with or without D-branes, taking into account the expansion of $F_3$, $H_3$ on the basis 
of $H^3_\ast(T^6, 4\mathbb{Z})$ for the case without discrete torsion and $H^3_\ast(T^6, 8\mathbb{Z})$ for the case with discrete torsion. Here, the $\Omega_3$ and $G_3$ are expanded with respect to 
a  basis of $H_-^3(T^6)$ for the orientifold with $O3/O7$-planes and 
and $H_+^3(T^6)$ for the orientifold with $O5/O9$-planes with $H_3$=0, 
where the number of untwisted moduli is counted by $h^{2,1}_-$ and 
$h^{2,1}_+$, respectively. (For the detailed discussions of toroidal 
orientifolds, see, e.g., Ref.~\cite{Blumenhagen:2006ci}.) 
Other toroidal orbifolds are discussed in Section \ref{sec:Ext}.

\subsection{$T^6/\mathbb{Z}_2$ and $T^6/(\mathbb{Z}_2\times \mathbb{Z}_2^\prime)$ 
with single complex structure modulus}

We first analyze the overall complex structure modulus, namely $\tau\equiv \tau_1=\tau_2=\tau_3$ and check whether the 4D CP is spontaneously broken or not. 
For the single complex structure modulus, the superpotential is simplified as
\begin{align}
    W =\left\{
\begin{array}{c}
     a^0 \tau^3  +3c S\tau^2 - 3 b \tau + d_0 S,\qquad (\gamma=0) \\
      -c^0S\tau^3 - 3a \tau^2 + 3 d S\tau - b_0,\qquad (\gamma=\pi)
\end{array}
\right.
,
\end{align}
where we define 
\begin{align}
a\equiv a^1=a^2=a^3,\quad b\equiv b_1=b_2=b_3,\quad  
c\equiv c^1=c^2=c^3,\quad d\equiv d_1=d_2=d_3.     
\end{align}
On the other hand, the K\"ahler potential is given by
\begin{align}
    K = -\ln (-i(S -\Bar{S})) -2\ln {\cal V} 
    -3\ln(-i(\tau-\Bar{\tau})).
\end{align}
As a consequence of the no-scale structure for the volume moduli, 
the moduli fields except for the volume moduli can be stabilized at the supersymmetric minimum $D_\tau W =D_S W=0$ with $D_IW=\partial_I W +(\partial_IK) W$~\cite{Honma:2019gzp},
\begin{itemize}
    \item $\gamma=0$ (mod $2\pi$)
\begin{align}
    {\rm Re}(\tau) ={\rm Re}(S) =0, \qquad
   {\rm Im}(\tau) =\left(-\frac{bd_0}{a^0c}\right)^{1/4},\quad
   {\rm Im}(S) =\left(-\frac{a^0(b)^3}{(c)^3d_0}\right)^{1/4},
\end{align}
    \item $\gamma=\pi$ (mod $2\pi$)
\begin{align}
    {\rm Re}(\tau) ={\rm Re}(S) =0, \qquad
   {\rm Im}(\tau) =\left(-\frac{b_0d}{ac^0}\right)^{1/4},\quad
   {\rm Im}(S) =\left(-\frac{(a)^3b_0}{c^0(d)^3}\right)^{1/4}.
\end{align}
\end{itemize}
As a result, the 4D CP is not spontaneously broken at the supersymmetric 
minimum for this simplest case. 
Hence, we move on to the case treating three independent complex structure moduli and search for the supersymmetric CP-breaking minimum in the next section.

\subsection{$T^6/\mathbb{Z}_2$ and $T^6/(\mathbb{Z}_2\times \mathbb{Z}_2^\prime)$ 
with three complex structure moduli}
\label{subsec:3_3}

We next explore the existence of CP-breaking minimum for the case with three complex structure moduli. 
We analytically discuss the spontaneous CP violation for $\gamma=0$ in Section~\ref{subsubsec_1} and $\gamma=\pi$ in Section~\ref{subsubsec_2}, 
respectively. 

\subsubsection{Odd polynomials ($\gamma=0$)}
\label{subsubsec_1}

In this section, we focus on the superpotential (\ref{eq:Wodd}) consisting of odd 
polynomials with respect to the moduli fields. 
The supersymmetric conditions $D_IW=0$ are still complicated equations 
with respect to the moduli fields, even for the CP invariant superpotential~(\ref{eq:Wodd}). 
To analytically solve the supersymmetric conditions, we simplify the effective action by redefining the 
moduli fields,
\begin{align}
    \tau_1&=\frac{c^1}{a^0} x_s\tau_1^\prime,\quad
    \tau_2=\frac{c^2}{a^0} x_s\tau_2^\prime,
\quad
    \tau_3=\frac{c^3}{a^0} x_s\tau_3^\prime,\quad
    S= x_s S^\prime.
\end{align}
According to the redefinition of moduli fields, the K\"ahelr potential 
and the superpotential are redefined as
\begin{align}
        K &= -\ln (-i(S -\Bar{S})) -2\ln {\cal V} 
    -\ln(i(\tau_1-\Bar{\tau}_1)(\tau_2-\Bar{\tau}_2)(\tau_3-\Bar{\tau}_3))    
\nonumber\\
&= -\ln (-i(S^\prime -\Bar{S}^\prime)) -2\ln {\cal V} 
    -\ln(i(\tau_1^\prime-\Bar{\tau}_1^\prime)(\tau_2^\prime-\Bar{\tau}_2^\prime)(\tau_3^\prime-\Bar{\tau}_3^\prime))
   -\ln (x_s^4 c^1c^2c^3/(a^0)^3),
\nonumber\\   
   W &= a^0 \tau_1\tau_2\tau_3  +c^1 S \tau_2\tau_3
    +c^2 S\tau_1\tau_3 +c^3 S\tau_1\tau_2
 - \sum_{i=1}^3 b_i \tau_i + d_0 S
  \nonumber\\ 
   &=\left\{ \tau_1^\prime\tau_2^\prime\tau_3^\prime  +S^\prime \tau_2^\prime\tau_3^\prime
    +S^\prime\tau_1^\prime\tau_3^\prime +S^\prime\tau_1^\prime\tau_2^\prime\right\}
 - \sum_{i=1}^3 b_i^\prime \tau_i^\prime + d_0^\prime S^\prime,
\end{align}
where we take
\begin{align}
    x_s =\left(\frac{(a^0)^2}{c^1c^2c^3}\right)^{1/3},\quad
    b_i^\prime =b_i \frac{c^i}{a^0}x_s,\quad
    d_0^\prime =d_0x_s.
\end{align}

In the following analysis, we omit prime symbols of fields unless specified otherwise.
Then, we analyze the following K\"ahler potential 
and superpotential:
\begin{align}
    K &= -\ln (-i(S -\Bar{S})) -2\ln {\cal V} 
    -\ln(i(\tau_1-\Bar{\tau}_1)(\tau_2-\Bar{\tau}_2)(\tau_3-\Bar{\tau}_3)),     
\nonumber\\
    W &= \tau_1\tau_2\tau_3  +S \tau_2\tau_3
    +S\tau_1\tau_3 +S\tau_1\tau_2
 - \sum_{i=1}^3 b_i \tau_i + d_0 S,
\label{eq:Woddred}
\end{align}
which is simplified version of the superpotential (\ref{eq:Wodd}) with 
$a^0=c^1=c^2=c^3=1$.

To analytically solve the supersymmetric conditions $D_IW=0$, we analyze the following equations 
equivalent to solve $D_IW=0$:
\begin{align}
    &{\rm Im}(\tau_1){\rm Re}(D_{\tau_1}W)\pm{\rm Im}(\tau_2){\rm Re}(D_{\tau_2}W)\pm{\rm Im}(\tau_3){\rm Re}(D_{\tau_3}W)-{\rm Im}(S){\rm Re}(D_{S}W)=0,
\nonumber\\
    &{\rm Im}(\tau_1){\rm Re}(D_{\tau_1}W)\mp{\rm Im}(\tau_2){\rm Re}(D_{\tau_2}W)\pm{\rm Im}(\tau_3){\rm Re}(D_{\tau_3}W)+{\rm Im}(S){\rm Re}(D_{S}W)=0,
\nonumber\\
    &{\rm Im}(\tau_1){\rm Im}(D_{\tau_1}W)\pm{\rm Im}(\tau_2){\rm Im}(D_{\tau_2}W)+{\rm Im}(\tau_3){\rm Im}(D_{\tau_3}W)\pm{\rm Im}(S){\rm Im}(D_{S}W)=0,
\nonumber\\
    &{\rm Im}(\tau_1){\rm Im}(D_{\tau_1}W)\mp{\rm Im}(\tau_2){\rm Im}(D_{\tau_2}W)-{\rm Im}(\tau_3){\rm Im}(D_{\tau_3}W)\pm{\rm Im}(S){\rm Im}(D_{S}W)=0,
\label{eq:SUSYeqsim}
\end{align}
where the double-sign corresponds. 
There is a CP-conserving solution, 
\begin{align}
    &{\rm Re}(\tau_1)={\rm Re}(\tau_2)={\rm Re}(\tau_3)={\rm Re}(S)=0,\quad
{\rm Im}\tau_1=\left(-\frac{b_2b_3d_0}{b_1} \right)^{1/4},
\nonumber\\
&{\rm Im}(\tau_2)=\left(-\frac{b_1b_3d_0}{b_2} \right)^{1/4},\quad
{\rm Im}(\tau_3)=\left(-\frac{b_1b_2d_0}{b_3} \right)^{1/4},\quad
{\rm Im}(S)=\left(-\frac{b_1b_2b_3}{d_0} \right)^{1/4},
\label{eq:Solodd}
\end{align}
at which all the moduli masses squared are positive.
In addition,  
we find that there exist five classes of solutions enumerated as follows: 

\begin{itemize}
\item Solution 1
\begin{align}
\frac{{\rm Re}(\tau_2)}{{\rm Re}(\tau_1)}&=
\frac{{\rm Im}(\tau_2)}{{\rm Im}(\tau_1)}=\sqrt{\frac{b_1}{b_2}},\quad
\frac{{\rm Re}(\tau_3)}{{\rm Re}(\tau_1)}= -
\frac{{\rm Im}(\tau_3)}{{\rm Im}(\tau_1)} = -\sqrt{\frac{b_1}{b_3}},
\nonumber\\
\frac{{\rm Re}(S)}{{\rm Re}(\tau_1)}&= \frac{{\rm Im}(S)}{{\rm Im}(\tau_1)}=\frac{\sqrt{b_1}}{\sqrt{b_3}-\sqrt{b_1}-\sqrt{b_2}},
\nonumber\\
{\rm Re}(\tau_1)&=\pm \frac{\sqrt{\sqrt{b_1} \left(-b_2 \sqrt{b_3}+\sqrt{b_2} b_3\right)-b_1
\left(\sqrt{b_2} \sqrt{b_3}+{\rm Im}(\tau_1)^2\right)}}{\sqrt{b_1}},
\end{align}
with $d_0= -\left(\sqrt{b_1}+\sqrt{b_2}-\sqrt{b_3}\right)^2$.

\item Solution 2
\begin{align}
\frac{{\rm Re}(\tau_2)}{{\rm Re}(\tau_1)}&= -
\frac{{\rm Im}(\tau_2)}{{\rm Im}(\tau_1)}= -\sqrt{\frac{b_1}{b_2}},\quad
\frac{{\rm Re}(\tau_3)}{{\rm Re}(\tau_1)}= -
\frac{{\rm Im}(\tau_3)}{{\rm Im}(\tau_1)} = -\sqrt{\frac{b_1}{b_3}},
\nonumber\\
\frac{{\rm Re}(S)}{{\rm Re}(\tau_1)}&= -\frac{{\rm Im}(S)}{{\rm Im}(\tau_1)}=\frac{\sqrt{b_1}}{\sqrt{b_3}-\sqrt{b_1}+\sqrt{b_2}},
\nonumber\\
{\rm Re}(\tau_1)&= \pm \frac{\sqrt{-\sqrt{b_1} \left(b_2 \sqrt{b_3}+\sqrt{b_2} b_3\right)+b_1
\left(\sqrt{b_2} \sqrt{b_3}-{\rm Im}(\tau_1)^2\right)}}{\sqrt{b_1}},
\end{align}
with $d_0= -\left(-\sqrt{b_1}+\sqrt{b_2}+\sqrt{b_3}\right)^2$.

\item Solution 3
\begin{align}
\frac{{\rm Re}(\tau_2)}{{\rm Re}(\tau_1)}&= -
\frac{{\rm Im}(\tau_2)}{{\rm Im}(\tau_1)}= -\sqrt{\frac{b_1}{b_2}},\quad
\frac{{\rm Re}(\tau_3)}{{\rm Re}(\tau_1)}= 
\frac{{\rm Im}(\tau_3)}{{\rm Im}(\tau_1)} = \sqrt{\frac{b_1}{b_3}},
\nonumber\\
\frac{{\rm Re}(S)}{{\rm Re}(\tau_1)}&= \frac{{\rm Im}(S)}{{\rm Im}(\tau_1)}=-\frac{\sqrt{b_1}}{\sqrt{b_3}+\sqrt{b_1}-\sqrt{b_2}},
\nonumber\\
{\rm Re}(\tau_1)&= \pm\frac{\sqrt{\sqrt{b_1} \left(b_2 \sqrt{b_3}-\sqrt{b_2} b_3\right)-b_1
\left(\sqrt{b_2} \sqrt{b_3}+{\rm Im}(\tau_1)^2\right)}}{\sqrt{b_1}},
\end{align}
with $d_0= -\left(\sqrt{b_1}-\sqrt{b_2}+\sqrt{b_3}\right)^2$.

\item Solution 4
\begin{align}
\frac{{\rm Re}(\tau_2)}{{\rm Re}(\tau_1)}&= 
\frac{{\rm Im}(\tau_2)}{{\rm Im}(\tau_1)}= \sqrt{\frac{b_1}{b_2}},\quad
\frac{{\rm Re}(\tau_3)}{{\rm Re}(\tau_1)}= 
\frac{{\rm Im}(\tau_3)}{{\rm Im}(\tau_1)} = \sqrt{\frac{b_1}{b_3}},
\nonumber\\
\frac{{\rm Re}(S)}{{\rm Re}(\tau_1)}&= -\frac{{\rm Im}(S)}{{\rm Im}(\tau_1)}=-\frac{\sqrt{b_1}}{\sqrt{b_3}+\sqrt{b_1}+\sqrt{b_2}},
\nonumber\\
{\rm Re}(\tau_1)&= \pm\frac{\sqrt{\sqrt{b_1} \sqrt{b_2} \left(\sqrt{b_1}+\sqrt{b_2}+\sqrt{b_3}\right)
\sqrt{b_3}-b_1 {\rm Im}(\tau_1)^2}}{\sqrt{b_1}},
\end{align}
with $d_0= -\left(\sqrt{b_1}+\sqrt{b_2}+\sqrt{b_3}\right)^2$.

\item Solution 5
\begin{align}
{\rm Im}(\tau_1)&= \sqrt{d_0},\quad
{\rm Re}(\tau_1)= 0,
\nonumber\\
{\rm Im}(\tau_2)&=\sqrt{d_0-{\rm Re}(\tau_2)^2},\quad
{\rm Im}(\tau_3)= \sqrt{d_0-{\rm Re}(\tau_3)^2},
\nonumber\\
{\rm Im}(S)&= \sqrt{d_0-{\rm Re}(S)^2},
\nonumber\\
{\rm Re}(S)&= \frac{-d_0\biggl[d_0{\rm Im}(\tau_3)^2-{\rm Im}(\tau_2)^2 \left({\rm Im}(\tau_3)^2+ \sqrt{\left(d_0-{\rm Im}(\tau_2)^2\right)
 \left(d_0-{\rm Im}(\tau_3)^2\right)}\right)\biggl]}{\sqrt{d_0-{\rm Im}(\tau_2)^2} \left(d_0{\rm Im}(\tau_2)^2 +(d_0-{\rm Im}(\tau_2)^2){\rm Im}(\tau_3)^2\right)},
\label{eq:solution5}
\end{align}
with $b_1=b_2=b_3= -d_0$.
\end{itemize}

In these five solutions, CP-breaking and -conserving minima are degenerate. 
That is, the above 5 solutions always include the flat directions in the moduli space of axio-dilaton 
and complex structure moduli, compared with the CP-conserving solution (\ref{eq:Solodd}). 
We will study the origin of the above flat directions in the flux vacua from the viewpoint of  modular symmetries in Section~\ref{subsec:CPvsMod}. 
The CP-breaking sources are then required to resolve the degeneracy between the CP-breaking and -conserving minima. 
Indeed, when we include the explicit CP-breaking term $b_0$ in the superpotential,
\begin{align}
    W = a^0 \tau_1\tau_2\tau_3  +c^1 S \tau_2\tau_3
    +c^2 S\tau_1\tau_3 +c^3 S\tau_1\tau_2
 - \sum_{i=1}^3 b_i \tau_i + d_0 S -b_0,
\end{align}
it is possible to resolve the degeneracy between the CP-breaking 
and -conserving minima. 
Note that the moduli fields are denoted by $\tau_i,S$ rather than $\tau_i^\prime, S^\prime$. 
For instance, under the following flux set
\begin{align}
   a^0=c^1=c^2=c^3=2,\quad b_0=d_0=-b_3=-2,\quad b_1=b_2=4 
\end{align}
leading to $n_{\rm flux}=24$, all the 
complex structure moduli and axio-dilaton are stabilized at CP-breaking minimum
\begin{align}\label{eq:explicit-CPV}
    {\rm Re}(\tau_1)&={\rm Re}(\tau_2)\simeq 0.127,\quad
    {\rm Re}(\tau_3)={\rm Re}(S)\simeq 0.258,    
\nonumber\\
    {\rm Im}(\tau_1)&={\rm Im}(\tau_2)\simeq 0.958,\quad
    {\rm Im}(\tau_3)={\rm Im}(S)\simeq 1.37,
\end{align}
with the masses squared of the moduli fields being positive
\begin{align}
{\cal V}^{-2}(19, 18, 9.3, 8.7, 8.2, 7.9, 4.0, 3.3)
\end{align}
in the descending order.

\subsubsection{Even polynomials ($\gamma=\pi$)}
\label{subsubsec_2}

Next, we analyze the superpotential (\ref{eq:Weven}) consisting 
of polynomials of even degree in the moduli fields. 
Similar to 
the analysis in Section~\ref{subsubsec_1}, we redefine the 
moduli fields
\begin{align}
    \tau_1&=\frac{a^1}{c^0x_s} \tau_1^\prime,\quad
    \tau_2=\frac{a^2}{c^0x_s} \tau_2^\prime,
\quad
    \tau_3=\frac{a^3}{c^0x_s} \tau_3^\prime,\quad
    S= x_s S^\prime,
\end{align}
simplifying the K\"ahelr potential 
and the superpotential as
\begin{align}
K &= -\ln (-i(S -\Bar{S})) -2\ln {\cal V} 
    -\ln(i(\tau_1-\Bar{\tau}_1)(\tau_2-\Bar{\tau}_2)(\tau_3-\Bar{\tau}_3))     
\nonumber\\
&= -\ln (-i(S^\prime -\Bar{S}^\prime)) -2\ln {\cal V} 
    -\ln(i(\tau_1^\prime-\Bar{\tau}_1^\prime)(\tau_2^\prime-\Bar{\tau}_2^\prime)(\tau_3^\prime-\Bar{\tau}_3^\prime))
   -\ln (a^1a^2a^3/(c^0)^3x_s^2),
\nonumber\\
   W &=  -c^0 S\tau_1\tau_2\tau_3 - a^1 \tau_2\tau_3
    - a^2\tau_1\tau_3 - a^3\tau_1\tau_2
+ \sum_{i=1}^3 d_i S \tau_i - b_0
  \nonumber\\ 
   &=-\left\{ S^\prime \tau_1^\prime\tau_2^\prime\tau_3^\prime  +
   \tau_2^\prime\tau_3^\prime
    +\tau_1^\prime\tau_3^\prime +\tau_1^\prime\tau_2^\prime\right\}
+ \sum_{i=1}^3 d_i^\prime S^\prime \tau_i^\prime - b^\prime_0,
\end{align}
where we define
\begin{align}
    x_s =\left(\frac{a^1a^2a^3}{(c^0)^2}\right)^{1/2},\quad
    d_i^\prime =d_i \frac{a^i}{c^0},\quad b_0^\prime=b_0.
\end{align}

In analogy with the analysis in the previous section, 
we omit prime symbols of fields unless specified otherwise.
Then, we analyze the following K\"ahler potential 
and superpotential:
\begin{align}
    K &= -\ln (-i(S -\Bar{S})) -2\ln {\cal V} 
    -\ln(i(\tau_1-\Bar{\tau}_1)(\tau_2-\Bar{\tau}_2)(\tau_3-\Bar{\tau}_3)),     
\nonumber\\
    W &= - S\tau_1\tau_2\tau_3 - \tau_2\tau_3
    - \tau_1\tau_3 - \tau_1\tau_2
+ \sum_{i=1}^3 d_i S \tau_i - b_0,
\label{eq:Wevenred}
\end{align}
which is the simplified version of the superpotential with $c^0=a^1=a^2=a^3=1$.

Following the analysis in the previous subsection, 
we solve the redefined supersymmetric equations (\ref{eq:SUSYeqsim}). 
It turns out that there exist a CP-conserving solution
\begin{align}
    &{\rm Re}(\tau_1)={\rm Re}(\tau_2)={\rm Re}(\tau_3)={\rm Re}(S)=0,\quad
{\rm Im}(\tau_1)=\left(-\frac{b_0d_2d_3}{d_1} \right)^{1/4},
\nonumber\\
&{\rm Im}(\tau_2)=\left(-\frac{b_0d_1d_3}{d_2} \right)^{1/4},\quad
{\rm Im}(\tau_3)=\left(-\frac{b_0d_1d_2}{d_3} \right)^{1/4},\quad
{\rm Im}(S)=\left(-\frac{b_0}{d_1d_2d_3} \right)^{1/4},
\label{eq:Soleven}
\end{align}
at which the masses squared of the moduli fields are positive. 
In addition, we find another solution:
\begin{itemize}
    \item  Solution 6
\begin{align}
    {\rm Im}(\tau_1) &=
    \sqrt{b_0-{\rm Re}(\tau_1)^2},
\nonumber\\
    {\rm Im}(\tau_2) &=\sqrt{b_0-{\rm Re}(\tau_2)^2},
\nonumber\\
    {\rm Im}(\tau_3) &=\sqrt{b_0-{\rm Re}(\tau_3)^2},
\nonumber\\
    {\rm Im}(S) &=\sqrt{\frac{1}{b_0}-{\rm  Re}(S)^2},
\nonumber\\
    {\rm Re}(\tau_1) &=-b_0\frac{{\rm Re}(\tau_2)+{\rm Re}(\tau_3)+b_0{\rm Re}(S)+{\rm Re}(\tau_2){\rm Re}(\tau_3){\rm Re}(S)}{b_0+{\rm Re}(\tau_2){\rm Re}(\tau_3)+b_0{\rm Re}(S)({\rm Re}(\tau_2)+{\rm Re}(\tau_3))},
\end{align}
with $d_{1}=d_2=d_3=-b_0$. 
\end{itemize}
As a result, there exist flat directions at the minimum due to the presence of unfixed moduli, and CP-breaking and -conserving minima are also degenerate. 
In the same way as in Section~\ref{subsubsec_1}, the CP-breaking 
sources are required to lift the flat directions.

We perform the numerical search to find the supersymmetric CP-breaking minima with 
masses squared of all the moduli fields being positive, but we cannot find such 
a solution as indicated by our analytical expressions.

\subsection{Extension to other toroidal orbifolds}
\label{sec:Ext}

So far, we have discussed the factorizable $T^6/\mathbb{Z}_2$ orientifold which is also applicable to $T^6/(\mathbb{Z}_2\times \mathbb{Z}_2')$ orientifold, taking into account the proper tadpole cancellation conditions. 
In these $T^6/\mathbb{Z}_2$ and $T^6/(\mathbb{Z}_2\times \mathbb{Z}_2')$ orientifolds, a largest number of flux quanta  (three-cycles) is allowed under the orbifold projections. 
In this section, we discuss other toroidal orbifolds where 
the possible three-cycles (\ref{eq:basis}) are restricted due to the orbifoldings. 

There exists single untwisted complex structure modulus on toroidal orbifolds including the factorizable $T_2/\mathbb{Z}_2$ such as 
$T^6/(\mathbb{Z}_2\times \mathbb{Z}_3) =  T^6/\mathbb{Z}_{6-II}$, $T^6/(\mathbb{Z}_2\times \mathbb{Z}_6)$,  $T^6/\mathbb{Z}_4$, $T^6/\mathbb{Z}_{8-II}$ and  $T^6/\mathbb{Z}_{12-II}$~\cite{Ibanez:1987pj,Font:1988mk,Katsuki:1989bf,Kobayashi:1991rp}, for the orbifolds preserving ${\cal N}=2$ supersymmetry, 
on which a part of three-form basis in Eq. (\ref{eq:basis}) remains under the orbifold 
projections~\cite{Lust:2005dy}.\footnote{
Whether there exists a single complex structure modulus or not depends on lattices to construct $T^6$.}
On the other hand, in other orbifolds, 
(untwisted) complex structure moduli are fixed at discrete values. 
When we label the single untwisted complex structure modulus $U$ 
on the above orbifolds, 
the 4D CP invariance is preserved for 
the following superpotential via the procedure in Section~\ref{sec:2_1},
\begin{itemize}
    \item $\gamma=0$ (mod $2\pi$)
\begin{align}
    W &= \xi_1 U + \xi_2 S,
\end{align}
    \item $\gamma=\pi$ (mod $2\pi$)
\begin{align}
    W &= \xi_3 S U + \xi_4,
\end{align}
\end{itemize}
where $\xi_{1,2,3,4}$ are integers constrained by the quantization 
conditions of $F_3, H_3$. 
Given the K\"ahler potential of $U$
\begin{align}
    K= -\ln (-i (U-\bar{U})),
\end{align}
the supersymmetric minimum is obtained by solving $D_UW=D_SW=0$,
\begin{itemize}
    \item $\gamma=0$ (mod $2\pi$)
\begin{align}
    {\rm Re}(U) =-\frac{\xi_2}{\xi_1}{\rm Re}(S),\quad
    {\rm Im}(U) =\frac{\xi_2}{\xi_1}{\rm Im}(S),    
\end{align}
    \item $\gamma=\pi$ (mod $2\pi$)
\begin{align}
    {\rm Re}(U) =-\frac{\xi_4}{\xi_3}\frac{{\rm Re}(S)}{{\rm Re}(S)^2 +{\rm Im}(S)^2},\quad
    {\rm Im}(U) =-\frac{\xi_4}{\xi_3}\frac{{\rm Im}(S)}{{\rm Re}(S)^2 +{\rm Im}(S)^2}.
\end{align}
\end{itemize}
At the minima, there exists a flat direction in the moduli space 
of axio-dilaton and complex structure modulus, even for the 
restricted orbifolds. 
Hence, we conclude that {\it Type IIB flux compactifications 
on toroidal orientifolds are not sufficient to realize the spontaneous 
CP violation.} 
Since flat directions generically appear at the degenerate CP-breaking 
and -conserving minima, it is required to consider the other CP-breaking sources which are discussed in Section~\ref{subsec:comment}.

\subsection{CP and modular symmetry}
\label{subsec:CPvsMod}

Before going to discuss the mechanism to lift the flat directions, 
we discuss a relation among the 4D CP, the modular symmetry, and flat directions, which appear in the above potential 
analysis.
In general, the presence of three-form fluxes breaks $SL(2,\mathbb{Z})_S$ and 
$SL(2,\mathbb{Z})_i$ modular symmetries associated with the axio-dilaton $S$ 
and three complex structure moduli $\tau_i$ in the low-energy effective action, 
respectively, but it 
is possible to preserve subgroups of the modular groups for a specific 
choice of three-form fluxes. (See for the classification of discrete 
modular symmetries, Ref.~\cite{Kobayashi:2020hoc}.) 

Let us focus on $T^6/\mathbb{Z}_2$ and $T^6/(\mathbb{Z}_2\times \mathbb{Z}^\prime_2)$ 
orientifolds with three complex structure moduli in Section~\ref{sec:3}. 
After redefining the moduli fields as in Eqs.~(\ref{eq:Woddred}) and (\ref{eq:Wevenred}), three-form fluxes are expanded as
\begin{itemize}
    \item $\gamma=0$ (mod $2\pi$)
\begin{align}
\frac{1}{l_s^2}F_3 &= \alpha_0+b^\prime_{i}\beta^{i}
=A^{(1)}_{ij}d\xi_1^id\xi_2^jdx^3+A^{(2)}_{ij}d\xi_1^id\xi_2^jdy^3,
\nonumber\\
\frac{1}{l_s^2}H_3 &= \alpha_{i} +d_0^\prime \beta^0
=A^{(3)}_{ij}d\xi_1^id\xi_2^jdx^3+A^{(4)}_{ij}d\xi_1^id\xi_2^jdy^3,
\label{eq:F3H3oddCP}
\end{align}
with $\xi_i =(y^i, x^i)^T$ and
\begin{align}
    A^{(1)}_{ij}=\left(
    \begin{array}{cc}
      -b^\prime_3 & 0  \\
        0  & 1
    \end{array}
     \right),
    A^{(2)}_{ij}=\left(
    \begin{array}{cc}
        0 & -b^\prime_2  \\
        -b^\prime_1  & 0
    \end{array}
     \right),
    A^{(3)}_{ij}=\left(
    \begin{array}{cc}
        0 & 1  \\
        1  & 0
    \end{array}
     \right),
    A^{(4)}_{ij}=\left(
    \begin{array}{cc}
      d^\prime_0 & 0  \\
        0  & 1
    \end{array}
\label{eq:Aodd}
     \right).
\end{align}
\item $\gamma=\pi$ (mod $2\pi$)
\begin{align}
\frac{1}{l_s^2}F_3 &= a^{i}\alpha_{i} +b_0 \beta^0
=B^{(1)}_{ij}d\xi_1^id\xi_2^jdx^3+B^{(2)}_{ij}d\xi_1^id\xi_2^jdy^3,
\nonumber\\
\frac{1}{l_s^2}H_3 &= c^0 \alpha_0 +d_{i}\beta^{i},
=B^{(3)}_{ij}d\xi_1^id\xi_2^jdx^3+B^{(4)}_{ij}d\xi_1^id\xi_2^jdy^3,
\label{eq:F3H3evenCP}
\end{align}
with 
\begin{align}
    B^{(1)}_{ij}=\left(
    \begin{array}{cc}
        0 & 1  \\
        1  & 0
    \end{array}
     \right),
    B^{(2)}_{ij}=\left(
    \begin{array}{cc}
      b^\prime_0 & 0  \\
        0  & 1
    \end{array}
     \right),
    B^{(3)}_{ij}=\left(
    \begin{array}{cc}
       -d^\prime_3 & 0  \\
        0  & 1
    \end{array}
     \right),
    B^{(4)}_{ij}=\left(
    \begin{array}{cc}
        0 & -d^\prime_2  \\
        -d^\prime_1  & 0
    \end{array}
     \right).
\label{eq:Beven}
\end{align}
\end{itemize}

When we consider the modular transformations on the tori $(T^2)_i$, 
the complex structure moduli as well as the coordinate of tori transform
\begin{align}
    \tau_i^\prime =R_i \tau_i,\quad
    \xi_i^\prime =(R_i^{-1})^{T}\xi_i,
\label{eq:xi}
\end{align}
with 
\begin{align}
    R_i = \left(
    \begin{array}{cc}
        p_i &  q_i\\
        r_i &  s_i 
    \end{array}
    \right)
    \in SL(2,\mathbb{Z})_i
\end{align}
satisfying $p_is_i -q_ir_i=1$. 
As discussed in detail in Ref.~\cite{Kobayashi:2020hoc}, the modular invariance of the 
effective action is realized when three-forms $F_3$ and $H_3$ 
themselves are invariant under the modular transformations.
Given the modular transformations on $(T^2)_1\times (T^2)_2$, 
the requirement for having the modular symmetries in the effective action leads to the modular transformations for the flux quanta in Eqs.~(\ref{eq:Aodd}) and (\ref{eq:Beven}),
\begin{align}
    A^{(m)}=R_1^{-1}A^{(m)}(R_2^{-1})^T,\quad
    B^{(m)}=R_1^{-1}B^{(m)}(R_2^{-1})^T,
\label{eq:req}
\end{align}
with $m=1,2,3,4$, where we use the transformations of $\xi_i$ in Eq.~(\ref{eq:xi}). Hence, the transformation matrices $R_{1,2}$ are constrained to satisfy 
the above relations for all $m$. 
From explicit expressions of $R_2$,
\begin{align}
    R_2&=(A^{(m)})^T (R_1^{-1})^T(A^{(m)})^{-1,T}\nonumber\\
       &=
    \biggl\{
    \left(
    \begin{array}{cc}
        s_1 & b_3^\prime r_1  \\
        \frac{q_1}{b_3^\prime}  & p_1
    \end{array}
     \right)
     \biggl|_{m=1},
     \left(
    \begin{array}{cc}
      p_1 & -\frac{b_1^\prime q_1}{b_2^\prime}  \\
    -\frac{b_2^\prime r_1}{b_1^\prime}& s_1
    \end{array}
     \right)
     \biggl|_{m=2},
     \left(
    \begin{array}{cc}
      p_1 & -q_1  \\
      -r_1& s_1
    \end{array}
     \right)
     \biggl|_{m=3},
    \left(
    \begin{array}{cc}
        s_1 & -d_0^\prime r_1  \\
        -\frac{q_1}{d_0^\prime}  & p_1
    \end{array}
     \right)
     \biggl|_{m=4}
     \biggl\},
\nonumber\\
    R_2&=(B^{(m)})^T (R_1^{-1})^T(B^{(m)})^{-1,T}\nonumber\\
       &=
    \biggl\{
    \left(
    \begin{array}{cc}
        p_1 & -q_1  \\
        -r_1  & s_1
    \end{array}
     \right)
     \biggl|_{m=1},
     \left(
    \begin{array}{cc}
        s_1 & d_3^\prime r_1  \\
        \frac{q_1}{d_3^\prime}  & p_1
    \end{array}
     \right)
     \biggl|_{m=2},
     \left(
    \begin{array}{cc}
        s_1 & -b_0^\prime r_1  \\
        -\frac{q_1}{b_0^\prime}  & p_1
    \end{array}
     \right)
     \biggl|_{m=3},
    \left(
    \begin{array}{cc}
      p_1 & -\frac{d_2^\prime q_1}{d_1^\prime}  \\
    -\frac{d_1^\prime r_1}{d_2^\prime}& s_1
    \end{array}
     \right)
     \biggl|_{m=4}
     \biggl\},
\end{align}
we find that only the following flux quanta 
\begin{align}
    &d_0^\prime=-b_3^\prime= 1,\quad b_1^\prime=b_2^\prime \quad (\gamma=0),
    \nonumber\\
   &d_3^\prime=-b_0^\prime= -1,\quad d_1^\prime=d_2^\prime \quad (\gamma=\pi)    
\label{eq:fluxT1T21}
\end{align}
with $q_1=r_1$, $p_1=s_1$ or
\begin{align}
    &d_0^\prime=-b_3^\prime= -1,\quad b_1^\prime=b_2^\prime \quad (\gamma=0),
    \nonumber\\
   &d_3^\prime=-b_0^\prime= 1,\quad d_1^\prime=d_2^\prime \quad (\gamma=\pi)   
\label{eq:fluxT1T22}
\end{align}
with $q_1=-r_1$, $p_1=s_1$ are solutions of Eq.~(\ref{eq:req}). 
Taking into account the condition det$(R_1)={\rm det}(R_2)=1$, the flux choice (\ref{eq:fluxT1T21}) 
corresponds to the trivial diagonal matrix for $R_2$ as well as $R_1$, 
whereas another choice (\ref{eq:fluxT1T22}) allows the $S$-transformation for the 
diagonal part of $SL(2,\mathbb{Z})_1\times SL(2,\mathbb{Z})_2$. 
We further analyze the modular transformation on $(T^2)_3$ which 
transforms the three-form fluxes $F_3$ and $H_3$ into
\begin{align}
\left(
    \begin{array}{c}
         F_3\\
         H_3 
    \end{array}
\right)
&=
\left(
\begin{array}{cc}
    {\cal C}_{ij}^{(2)}d\xi_1^id\xi_2^j &  
    {\cal C}_{ij}^{(1)}d\xi_1^id\xi_2^j\\
    {\cal C}_{ij}^{(4)}d\xi_1^id\xi_2^j &  
    {\cal C}_{ij}^{(3)}d\xi_1^id\xi_2^j
\end{array}
\right)
\left(
    \begin{array}{c}
         y^3\\
         x^3 
    \end{array}
\right)
\nonumber\\
&\rightarrow
\left\{
\left(
\begin{array}{cc}
    {\cal C}_{ij}^{(2)}d\xi_1^id\xi_2^j &  
    {\cal C}_{ij}^{(1)}d\xi_1^id\xi_2^j\\
    {\cal C}_{ij}^{(4)}d\xi_1^id\xi_2^j &  
    {\cal C}_{ij}^{(3)}d\xi_1^id\xi_2^j
\end{array}
\right)
(R_3^{-1})^T
\right\}
\left(
    \begin{array}{c}
         y^3\\
         x^3 
    \end{array}
\right)
,
\end{align}
with ${\cal C}_{ij}^{(m)}=\{A_{ij}^{(m)},B_{ij}^{(m)}\}$, 
which is not invariant under $SL(2,\mathbb{Z})_3$ itself for a generic choice of fluxes. 
Indeed, $S$- and $T$-transformations of $SL(2,\mathbb{Z})_3$ change the three-form fluxes 
$(F_3, H_3)$ themselves, namely
\begin{align}
\left(
    \begin{array}{c}
         F_3\\
         H_3 
    \end{array}
\right)
&\rightarrow
\left\{
\begin{array}{c}
\left(
\begin{array}{cc}
    -{\cal C}_{ij}^{(1)}d\xi_1^id\xi_2^j &  
    {\cal C}_{ij}^{(2)}d\xi_1^id\xi_2^j\\
    -{\cal C}_{ij}^{(3)}d\xi_1^id\xi_2^j &  
    {\cal C}_{ij}^{(4)}d\xi_1^id\xi_2^j
\end{array}
\right)
\left(
    \begin{array}{c}
         y^3\\
         x^3 
    \end{array}
\right),
\quad
{\rm with}\,\,R_3=
\left(
\begin{array}{cc}
    0 &  1\\
    -1 & 0
\end{array}
\right)
\\
\left(
\begin{array}{cc}
    -{\cal C}_{ij}^{(1)}d\xi_1^id\xi_2^j +{\cal C}_{ij}^{(2)}d\xi_1^id\xi_2^j &  
    {\cal C}_{ij}^{(1)}d\xi_1^id\xi_2^j\\
    -{\cal C}_{ij}^{(3)}d\xi_1^id\xi_2^j +{\cal C}_{ij}^{(4)}d\xi_1^id\xi_2^j&  
    {\cal C}_{ij}^{(3)}d\xi_1^id\xi_2^j
\end{array}
\right)
\left(
    \begin{array}{c}
         y^3\\
         x^3 
    \end{array}
\right),
\quad
{\rm with}\,\,R_3=
\left(
\begin{array}{cc}
    1 &  1\\
    0 & 1
\end{array}
\right)
\end{array}
\right.
,
\end{align}
which is not identical to the original $(F_3, H_3)$. 

However, there is a chance to keep the (discrete) modular symmetry on 
$(T^2)_3$, identifying the modular transformation $R_3$ of $SL(2,\mathbb{Z})_3$ 
with $SL(2,\mathbb{Z})_S$ modular transformation $R$ associated with the axio-dilaton. 
Note that $SL(2,\mathbb{Z})_S$ itself always exists in the effective action, where the axio-dilaton as well as the three-form fluxes 
transform
\begin{align}
    S\rightarrow R(S),\quad
    \left(
    \begin{array}{c}
         F_3\\
         H_3 
    \end{array}
\right)
\rightarrow R
\left(
    \begin{array}{c}
         F_3\\
         H_3 
    \end{array}
\right)
,
\end{align}
for $R\in SL(2,\mathbb{Z})_S$. 

For the following choice of fluxes with 
\begin{align}
    d_0^\prime= \mp b_3^\prime,\quad b_1^\prime=b_2^\prime= \pm 1,\quad (\gamma=0),
    \nonumber\\
    d_3^\prime= \mp b_0^\prime,\quad d_1^\prime=d_2^\prime= \pm 1,\quad (\gamma=\pi),    
\label{eq:fluxT3}
\end{align}
the three-form fluxes $(F_3,H_3)$ after performing the $S$-transformation of $SL(2,\mathbb{Z})_3$
\begin{align}
\left(
    \begin{array}{c}
         F_3\\
         H_3 
    \end{array}
\right)
&\rightarrow
R R^{-1}
\left(
\begin{array}{cc}
    -{\cal C}_{ij}^{(1)}d\xi_1^id\xi_2^j &  
     {\cal C}_{ij}^{(2)}d\xi_1^id\xi_2^j\\
    -{\cal C}_{ij}^{(3)}d\xi_1^id\xi_2^j &  
     {\cal C}_{ij}^{(4)}d\xi_1^id\xi_2^j
\end{array}
\right)
\left(
    \begin{array}{c}
         y^3\\
         x^3 
    \end{array}
\right)
=
R
\left(
    \begin{array}{c}
         F_3\\
         H_3 
    \end{array}
\right)
\end{align}
with
\begin{align}
    R =
\left(
\begin{array}{cc}
    0 &  \mp 1\\
    \pm 1 & 0
\end{array}
\right)\
,
\label{eq:RT3}
\end{align}
allow us to identify $R$ with the element of $SL(2,\mathbb{Z})_S$. 
Here, the double-sign corresponds in Eqs.~(\ref{eq:fluxT3}) and (\ref{eq:RT3}). 
Hence, the $S$-transformation of diagonal $SL(2,\mathbb{Z})_3\times SL(2,\mathbb{Z})_S$ exists in the potential for a particular choice of fluxes, but it is difficult to 
realize the $T$-transformation of $SL(2,\mathbb{Z})_3$ even when 
the modular transformation of the axio-dilaton is taken into account. 
The spontaneous symmetry breaking of these modular symmetries 
will be analyzed in the actual flux vacua later. 
Although we choose a particular basis of the three-form fluxes (\ref{eq:F3H3oddCP}) 
and (\ref{eq:F3H3evenCP}), namely $\{\xi_1, \xi_2\}$, but it is possible to consider  other 
bases such as $\{\xi_1, \xi_3 \}$ and $\{\xi_2, \xi_3\}$. 

In the following, we analyze the modular symmetry of the flux vacua which 
is the remnant of modular symmetry in the effective action discussed so far. 
Remarkably, all the solutions we found in Section~\ref{sec:3} satisfy the relation
\begin{align}
    {\rm Re}(\phi_m)^2+{\rm Im}(\phi_m)^2=1,
\label{eq:Abs1}
\end{align}
with $\phi_m$ being the rescaled moduli for the solutions:
\begin{itemize}
    \item Solution 1
\begin{align}
    \phi_1=\left(\frac{b_1^\prime}{b_2^\prime b_3^\prime}\right)^{1/4}\frac{\tau_1^\prime}{p_1},\quad
    \phi_2=\left(\frac{b_2^\prime}{b_1^\prime b_3^\prime}\right)^{1/4}\frac{\tau_2^\prime}{p_1},\quad
    \phi_3=\left(\frac{b_3^\prime}{b_1^\prime b_2^\prime}\right)^{1/4}\frac{\tau_3^\prime}{p_1},\quad
    \phi_4=\frac{p_1S^\prime}{(b^\prime_1b^\prime_2b^\prime_3)^{1/4}},
\end{align}
with $p_1\equiv (-\sqrt{b^\prime_1}-\sqrt{b^\prime_2}+\sqrt{b^\prime_3})^{1/2}$.
    \item Solution 2
\begin{align}
    \phi_1=\left(\frac{b^\prime_1}{b^\prime_2b^\prime_3}\right)^{1/4}\frac{\tau^\prime_1}{p_2},\quad
    \phi_2=\left(\frac{b^\prime_2}{b^\prime_1b^\prime_3}\right)^{1/4}\frac{\tau^\prime_2}{p_2},\quad
    \phi_3=\left(\frac{b^\prime_3}{b^\prime_1b^\prime_2}\right)^{1/4}\frac{\tau^\prime_3}{p_2},\quad
    \phi_4=\frac{p_2S^\prime}{(b^\prime_1b^\prime_2b^\prime_3)^{1/4}},
\end{align}
with $p_2\equiv (\sqrt{b^\prime_1}-\sqrt{b^\prime_2}-\sqrt{b^\prime_3})^{1/2}$.
    \item Solution 3
\begin{align}
    \phi_1=\left(\frac{b^\prime_1}{b^\prime_2b^\prime_3}\right)^{1/4}\frac{\tau_1^\prime}{p_3},\quad
    \phi_2=\left(\frac{b^\prime_2}{b^\prime_1b^\prime_3}\right)^{1/4}\frac{\tau_2^\prime}{p_3},\quad
    \phi_3=\left(\frac{b^\prime_3}{b^\prime_1b^\prime_2}\right)^{1/4}\frac{\tau_3^\prime}{p_3},\quad
    \phi_4=\frac{p_3S^\prime}{(b^\prime_1b^\prime_2b^\prime_3)^{1/4}},
\end{align}
with $p_3\equiv (-\sqrt{b^\prime_1}+\sqrt{b^\prime_2}-\sqrt{b^\prime_3})^{1/2}$.
    \item Solution 4
\begin{align}
    \phi_1=\left(\frac{b^\prime_1}{b^\prime_2b^\prime_3}\right)^{1/4}\frac{\tau_1^\prime}{p_4},\quad
    \phi_2=\left(\frac{b^\prime_2}{b^\prime_1b^\prime_3}\right)^{1/4}\frac{\tau_2^\prime}{p_4},\quad
    \phi_3=\left(\frac{b^\prime_3}{b^\prime_1b^\prime_2}\right)^{1/4}\frac{\tau_3^\prime}{p_4},\quad
    \phi_4=\frac{p_4S^\prime}{(b^\prime_1b^\prime_2b^\prime_3)^{1/4}},
\end{align}
with $p_4\equiv (\sqrt{b^\prime_1}+\sqrt{b^\prime_2}+\sqrt{b^\prime_3})^{1/2}$.
    \item Solution 5
\begin{align}
    \phi_1=\frac{\tau_1^\prime}{\sqrt{d^\prime_0}},\quad
    \phi_2=\frac{\tau_2^\prime}{\sqrt{d^\prime_0}},\quad
    \phi_3=\frac{\tau_3^\prime}{\sqrt{d^\prime_0}},\quad
    \phi_4=\frac{S^\prime}{\sqrt{d^\prime_0}}.
\end{align}
    \item Solution 6
\begin{align}
    \phi_1=\frac{\tau_1^\prime}{\sqrt{b_0^\prime}},\quad
    \phi_2=\frac{\tau_2^\prime}{\sqrt{b_0^\prime}},\quad
    \phi_3=\frac{\tau_3^\prime}{\sqrt{b_0^\prime}},\quad
    \phi_4=\sqrt{b_0^\prime}S^\prime.
\end{align}
\end{itemize}
Here we have explicitly written prime symbols in the solutions.

In this way, the flux vacua are distributed along the circumference of the unit circle as in Eq.~(\ref{eq:Abs1}) for the rescaled moduli. 
It means that when the rescaled moduli $\phi_m$ are equal to the 
original unrescaled moduli fields, some of $S$-transformations in $\Pi_{i=1}^3 SL(2,\mathbb{Z})_i\times SL(2,\mathbb{Z})_S$ modular groups, \begin{align}
\phi_m \rightarrow -(\phi_m)^{-1} =-\frac{\bar{\phi}_m}{|\phi_m|},
\end{align}
with $|\phi_m|=1$ are manifest in the obtained flux vacua 
for a particular choice of fluxes. 
Furthermore, $S$-transformations in the flux vacua with $|\tau_i|=1$ correspond to the CP transformations $\tau_i\rightarrow -\bar{\tau}_i$ as discussed in Section~\ref{sec:2}.
The presence of CP transformation at the flux vacua also supports the existence of unbroken 4D CP. 
Note that the solutions found in Section~\ref{subsec:3_3} are parametrized by more generic fluxes, 
compared with the flux vacua having the (discrete) modular symmetry. 
To clarify this point, we show the solution 5 with the specific flux quanta $d_0^\prime=-b_1^\prime=-b_2^\prime=-b_3^\prime=1$ which lead to the following superpotential
\begin{align}
       W&=\left\{ \tau_1^\prime\tau_2^\prime\tau_3^\prime  +S^\prime \tau_2^\prime\tau_3^\prime
    +S^\prime\tau_1^\prime\tau_3^\prime +S^\prime\tau_1^\prime\tau_2^\prime\right\}
 +\sum_{i=1}^3 \tau_i^\prime + S^\prime.
\end{align}
The explicit form of the solution 5 is given by
\begin{align}
    {\rm Re}(\phi_m)^2+{\rm Im}(\phi_m)^2=d_0^\prime=1,
\label{eq:abs}
\end{align}
where $\phi_m =\{\tau_i^\prime, S^\prime\}$ are distributed along the circumference of the unit 
circle. However, the above solutions do not satisfy the condition (\ref{eq:fluxT3}), meaning that only the $S$-transformation of the diagonal $SL(2,\mathbb{Z)}_1\times SL(2,\mathbb{Z})_2$ remains in the flux vacua. 
Thus, the flat directions are remnant of the modular symmetries in rescaled moduli.
Furthermore, the 4D CP can be embedded into modular symmetry in the background of three-form fluxes  such that 
the flat directions correspond to the unit circle in the unrescaled moduli.
The other choices of three-form fluxes give rise to the circle with  different radii and the 4D CP is not identified 
with the $S$-transformation.

\subsection{Comments on the CP-breaking sources}
\label{subsec:comment}

In this section, we comment on sources of CP violation 
to resolve the degeneracy between CP-breaking and -conserving 
minima. 

So far, we have focused on the complex structure moduli and 
dilaton, but the unfixed K\"ahler moduli whose axions determine 
the size of $\theta$ term still play an important role of CP violation 
and affect the dynamics of the other moduli. 
To stabilize the K\"ahler moduli, we introduce the non-perturbative 
effects to the superpotential
\begin{align}
    W = \gamma(\tau, S)e^{-f T},
\end{align}
where $T$ denotes the K\"ahler moduli appearing in the gaue kinetic 
function (\ref{eq:gaugekin}). 
Here, $\gamma(\tau, S)$ is supposed to the one-loop determinant of the D3-brane instanton effects or one-loop corrections to the gauge couplings on the D7-branes 
where the gauginos condensate. 
Although the lifting of flat directions in the CP-breaking and -conserving minima 
is highly dependent on the functional form of $\gamma(\tau,S)$ as well as the mechanism to uplift the anti-de Sitter vacuum obtained after having fixed all the moduli to de Sitter one, it might lead to 
the realization of spontaneous CP violation even when $\gamma(\tau, S)$ 
is a CP-invariant function. 

Another possibility to realize the spontaneous CP violation is to 
consider Calabi-Yau threefolds or blowing up the orbifolds as an extension of toroidal oribifolds. 
It is interesting to discuss the Calabi-Yau compactifications with 
three-form fluxes, but the number of complex structure moduli is typically of ${\cal O}(100)$ which is hard to analyze the CP-breaking minima analytically. 
Also, the introduction of “geometric fluxes" as well as the non-geometric fluxes 
will be useful to realize the spontaneous CP violation, because such fluxes 
induce the couplings between the K\"ahler moduli and complex structure moduli~\cite{Gurrieri:2002wz,Kachru:2002sk,Shelton:2005cf}. 
We leave the detailed study of these possibilities for future work.

\section{Conclusions}
\label{sec:con}

In this paper, we systematically investigated whether the spontaneous CP violation can be realized in Type IIB flux compactifications on toroidal orientifolds, which allow the moduli stabilization in a controlled way. 

Similar to the heterotic string theory, 4D CP is regarded 
as a discrete gauge symmetry embedded into the 10D proper Lorentz 
symmetry. 
To ensure the presence of 4D CP in the potential, the three-cycles three-form fluxes turn on are restricted to even or odd polynomials with respect to the complex structure moduli 
associated with tori and the axio-dilaton. 
Our detailed analysis shows that there exist flat directions at the degenerate CP-breaking and -conserving vacua for a generic choice of fluxes. Hence, the flux compactifications are insufficient to realize the spontaneous CP violation in Type IIB toroidal orientifolds. 
The statement also holds for the heterotic string theory as well as Type IIA string theory 
on toroidal orbifolds with fluxes, because the functional form of the flux-induced superpotential 
is restricted one of Type IIB flux compactifications. 
These flat directions are remnants of the modular $S$-transformation for rescaled moduli. 
Furthermore, the 4D CP is embedded into the modular symmetries 
in the effective action for a particular choice of fluxes and 
such an approach to unify CP and modular symmetries is closely related to the recent 
discussion in the heterotic orbifold context~\cite{Nilles:2018wex,Dent:2001cc,Baur:2019kwi,Baur:2019iai}, 
but in the different context.

To resolve the degeneracy between the CP-breaking and -conserving vacua, it is required to extend our system to introduce non-perturbative effects with respect to the K\"ahler moduli and/or (non-)geometric fluxes in the superpotential. Calabi-Yau compactifications as well as the blown up orbifold compactifications will give a crucial role of lifting the 
flat directions. 
We would report this interesting work in the future.

Recently, modular flavor symmetries were studied to derive realistic quark and lepton mass matrices. 
(See e.g. \cite{Feruglio:2017spp}.)
Although the modular symmetry is completely broken at generic value of $\tau$, 
its subgroups remain at fixed points, e.g. $Z_3$ symmetry at $\tau = e^{i\pi/3}$, $Z_2$ symmetry at $\tau=i$ and   
$Z_2$ symmetry at $\tau = i \infty$.
In this sense, moduli values at and near these fixed points are interesting. 
(See e.g. \cite{deAnda:2018ecu}.)
For example, realistic quark and lepton mass matrices are obtained around $\tau=i$ \cite{Okada:2019uoy}.
Our analysis can lead to such values, i.e. $\tau_1=i$ for $d_0=1$ in Solution 5 and $\tau_{1,2}=0.127 + 0.958 i$ 
in Eq.~(\ref{eq:explicit-CPV}) as well as $\tau \sim i \infty$ for sufficiently large $d_0$ in Solution 5.
Our result would also be important from this viewpoint.

\subsection*{Acknowledgements}

T. K. was supported in part by MEXT KAKENHI Grant Number JP19H04605. 
H. O. was supported in part by JSPS KAKENHI Grant Numbers JP19J00664 and JP20K14477.

\end{document}